\documentclass[aps,twocolumn,superscriptaddress]{revtex4}
\usepackage{amsmath}
\usepackage{amsfonts}
\usepackage{mathrsfs}
\usepackage{graphicx}
\usepackage{times}
\usepackage{color}
\usepackage[normalem]{ulem}

\newcommand\redout{\bgroup\markoverwith{\textcolor{red}{\rule[.5ex]{2pt}{1pt}}}\ULon}

\def\be{\begin{equation}}
\def\ee{\end{equation}}
\def\bea{\begin{eqnarray}}
\def\eea{\end{eqnarray}}

\begin{document}

\title{Topological semimetal and superfluid of $s$-wave interacting fermionic atoms in an orbital optical lattice}

\author{Maksims Arzamasovs}
\affiliation{MOE Key Laboratory for Nonequilibrium Synthesis and Modulation of Condensed Matter,Shaanxi Province Key Laboratory of Quantum Information and Quantum Optoelectronic Devices, School of Physics, Xi'an Jiaotong University, Xi'an 710049, China}

\author{Shuai Li}
\affiliation{MOE Key Laboratory for Nonequilibrium Synthesis and Modulation of Condensed Matter,Shaanxi Province Key Laboratory of Quantum Information and Quantum Optoelectronic Devices, School of Physics, Xi'an Jiaotong University, Xi'an 710049, China}

\author{W. Vincent Liu}
\email{wvliu@pitt.edu}
\affiliation{Department of Physics and Astronomy, University of Pittsburgh, Pittsburgh PA 15260, USA}
\affiliation{Wilczek Quantum Center, School of Physics and Astronomy and T. D. Lee Institute, Shanghai Jiao Tong University, Shanghai 200240, China}		
\affiliation{Shanghai Research Center for Quantum Sciences, Shanghai 201315, China}

\author{Bo Liu}
\email{liubophy@gmail.com}
\affiliation{MOE Key Laboratory for Nonequilibrium Synthesis and Modulation of Condensed Matter,Shaanxi Province Key Laboratory of Quantum Information and Quantum Optoelectronic Devices, School of Physics, Xi'an Jiaotong University, Xi'an 710049, China}

\begin{abstract}
   {Recent advanced experimental implementations of optical lattices
    with highly tunable geometry open up new regimes for quantum many-body
    states of matter that previously had not been accessible. Here we introduce
    a symmetry-based method of utilizing the geometry of optical lattice to
    systematically control topologically non-trivial orbital hybridization. Such
    an orbital mixing leads to an unexpected and yet robust topological semimetal
    at single{-}particle level for a gas of fermionic atoms. When considering
    $s$-wave attractive interaction between atoms as for instance tuned by
    Feshbach resonance, topological superfluid state 
    with high Chern
    number is unveiled in the presence of {on-site} rotation. This state
    supports chiral edge excitations, manifesting its topological nature.  An
    experimental realization scheme is designed, which introduces a systematic
    way {of achieving} a new universality class (such as Chern number of $2$) of
    orbital-hybridized topological phases beyond geometrically standard optical lattices.  }

\end{abstract}

\maketitle
Pursuit of topological phases of matter is one of the central {thrusts} in
condensed matter physics since {the discovery
of superfluid $^{3}$He chiral A phase~\cite{2003_Volovik_book} and quantum Hall
effect~\cite{RevModPhys1,RevModPhys2}}. The concept of topology not only plays a key role in a variety of
exotic quantum phenomena, such as topological insulators, chiral
superconductors or Weyl semimetals, but is also closely related to fundamental
physics, i.e., to distinguish new phases of matter that cannot be characterized by broken symmetries. It has thus explosively triggered a tremendous amount of efforts
in both theoretical and experimental studies in {various} solid state
materials. Besides the continuously growing efforts in solids, there has
been a great interest {in} simulating topological matter {with} ultracold {gases}.
Such highly controllable atomic systems will not only provide a versatile
tool for simulating electronic systems, but also supply new {possibilities} to
study new phenomena with no {counterpart} in solids. Recent experimental
advances {in creating} tunable spin-orbit coupling by using the Raman scheme
provide unprecedented opportunities for the study of topological matter in
ultracold gases~\cite{GalitskiiSpielman,2011_RevModPhys,LinJimenezGarciaSpielman,WuZhangSunXuWangJiDengChenLiuPan,2012_Cheuk_PhysRevLett, 2012_Wang_PhysRevLett,2014_Jotzu_Nature,DucaLiReitterBlochEtal,2013_Aidelsburger_PhysRevLett, 2013_Miyake_PhysRevLett,2013_cheng_Natphys,2012_huizhai_IJMP}. Moreover, the hybridization of orbital bands with different parity has opened a completely different avenue {to emulate} spin-orbit coupling or artificial gauge fields in general for cold atoms~\cite{2014_Qizhou_PhysRevA,2014_huizhai_PhysRevA,2014_cooper_PhysRevA,2007_Bloch_PRL,2011_Hemmerich_NatPhys,2012_Sengstock_Natphys,2015_Hemmerich_PhysRevLett}, yielding various interesting quantum states of matter ~\cite{2012_Kaisun_naturephy,2013_xiaopeng_Natcommun,2014_Bo_arxiv,2018_Bo_PhysRevLett,2008_erhai_PhysRevLett,2008_congjun_PhysRevLett,
2008_2_congjun_PhysRevLett}. However, the way to systematically engineer the orbital-hybridized topological phases of matter in static optical lattices remains unclear and stands as an obstacle of a substantial effort.

\begin{figure}[t]
\includegraphics[width=9cm]{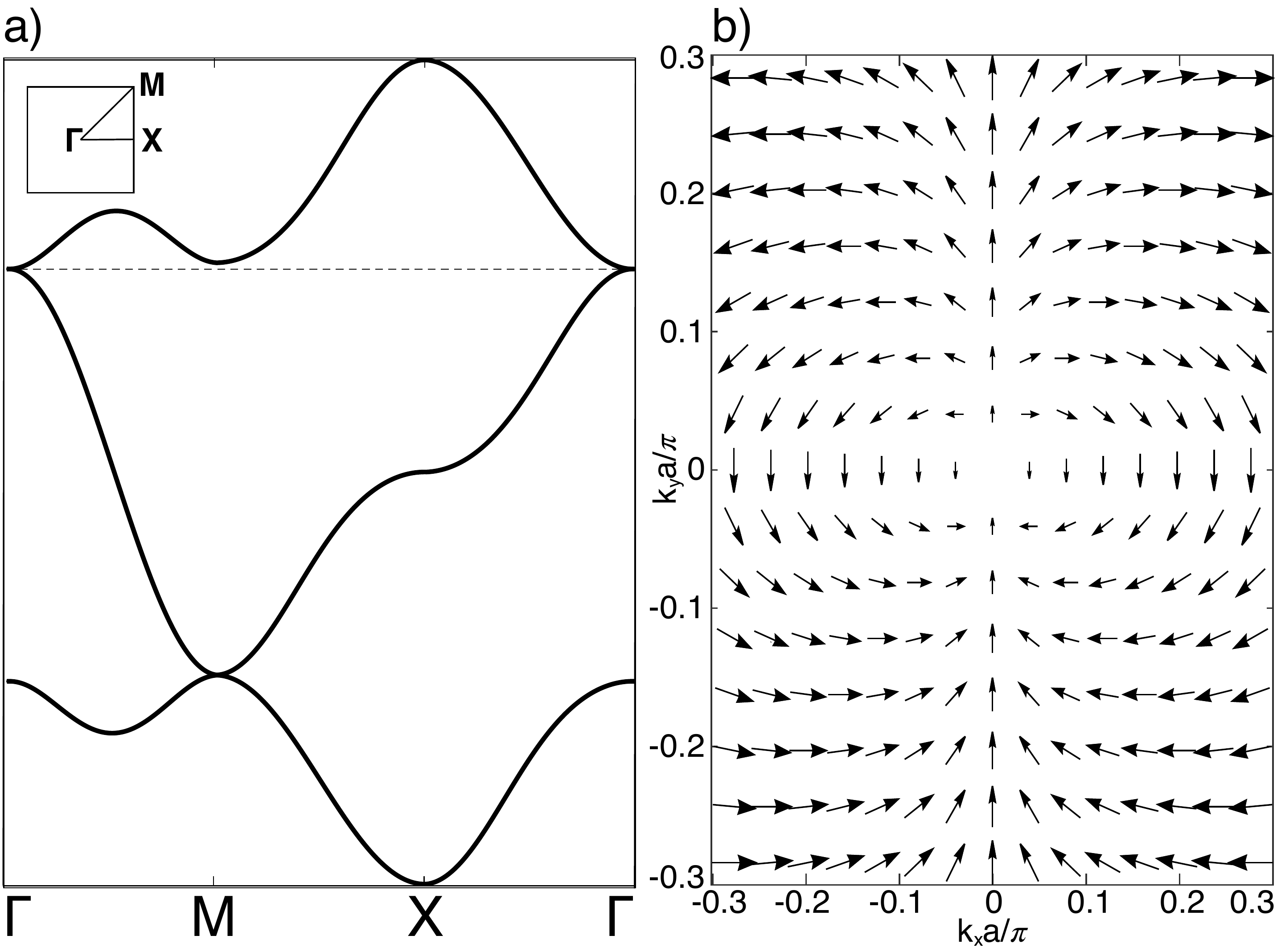}
\caption{(a) Single-particle energy spectrum of the  tight-binding  model
in Eq.~\eqref{eq:LatticeHamiltonian}. We show the band structure
along a contour in momentum space demonstrated in the inset of Fig.~\ref{fig:QuiverPlot}(a). The band degeneracy point appears
at $\Gamma$ with the parabolic dispersion. (b) The topological nature of
band degeneracy point. The planar vector $\mathbf{h}$ defined in Eq.~\eqref{eq:TwoVectorHVortex} forms a topological defect in $(k_x,k_y)$-plane,
which is a vortex in the momentum space with winding number $2$.
Other parameters are chosen as $t_m=0.6t_{\parallel}$ and $t_z=t_{\perp}=0.3t_{\parallel}$.}
\label{fig:QuiverPlot}
\end{figure}

{In this work}, we report the discovery of a new mechanism to explore various
unexpected orbital-hybridized topological phases. To demonstrate
this scheme, a concrete model of ultracold fermionic atoms in an optical
lattice {is} introduced below. The key idea here is to introduce the
symmetry-based method of systematically controlling the non-trivial
hybridization between degenerate orbitals of ultracold atoms in optical
lattices. Surprisingly, we unveil that the interplay {between manipulating} the
inversion symmetry of the system and orbital degeneracy of cold atoms in
optical lattices can lead to a novel type of topologically non-trivial
semimetal, where neither Raman-induced spin-orbit coupling nor other
artificial gauge field is required. When further considering attraction between
fermionic atoms, an s-wave interaction induced topological superfluid with
high Chern number emerges in the presence of on-site rotation. This idea
is motivated by the recent
experimental advances in manipulating higher orbital bands in optical
lattices, such as the breakthrough observation of long-lived $p$-band
bosonic atoms in a checkerboard lattice~\cite{2007_Bloch_PRL,2011_Hemmerich_NatPhys,2012_Sengstock_Natphys,
2015_Hemmerich_PhysRevLett}. It provides unprecedented
opportunities to investigate quantum many-body phases with orbital degrees
of freedom~\cite{2014_Bo_arxiv,2016_Bo_PhysRevA,2016_Bo_PhysRevAI,2013_xiaopeng_Natcommun,2010_Zixu_PhysRevA,
2011_Zicai_PRA,2012_Zicai_PhysRevB,2008_erhai_PhysRevLett,2008_congjun_PhysRevLett,2008_2_congjun_PhysRevLett,
2015_Erhai_PhysRevLett,2013_Pinheiro_PhysRevLett,2012_xiaopeng_PhysRevLett}. As we shall show with the model below, the symmetry-based
manipulation of non-trivial hybridization between degenerate orbitals can
lead to other unexpected results.

\textit{Orbital-hybridized topological semimetal $\raisebox{0.01mm}{---}$}%
Let us consider an ultracold fermionic gas loaded in a 2D optical lattice
which can be realized from a strongly anisotropic 3D optical lattice.
Specifically, the lattice potential can be expressed as $V_{\mathrm{OL}}({%
\mathbf{r}})=-V_x \cos^2(k_{Lx}x)-V_y \cos^2(k_{Ly}y) -V_z \cos^2(k_{Lz}z)$
with the wavevectors of the laser fields $k_{Lx}$, $k_{Ly}$ and $k_{Lz}$.
The corresponding lattice constants are $a_x=\pi/k_{Lx}$, $%
a_y=\pi/k_{Ly}$, $a_z=\pi/k_{Lz}$ along the $x$, $y$ and $z$ directions,
respectively. Here we focus on the case with lattice depth $%
V_{z}>>V_{x}=V_{y}$ and the system dynamically behaves as a 2D system. In
the deep lattice limit, the lattice potential at each site can be
approximated by a harmonic oscillator. Under this approximation, the
requirement of $V_xk^2_{Lx}=V_yk^2_{Ly}=V_zk^2_{Lz}$ would keep the local
SO(3) rotation symmetry at each lattice site, which guarantees the
three-fold degeneracy of $p$-orbitals {locally}. Another key ingredient of our
scheme is {to include} a {magnetic field gradient} along the $z$-direction.
The single-particle physics can thus be captured by the following Hamiltonian, $%
H_0=-\frac{\hbar^2}{2m}\nabla^2+V_{\mathrm{OL}}(\mathbf{r})-\mathbf{F}\cdot
\mathbf{r}$, where $\mathbf{F}=-J\nabla_z B$ is the force applied to the
atom with spin magnetic moment $J$ . Here we want to emphasize the key role
of the external {magnetic field gradient} in our proposal. First, it breaks
the inversion symmetry along the $z$-direction and thus induces the
non-trivial hybridization between the degenerate orbitals, i.e., $p_x$ and $%
p_z$, $p_y$ and $p_z$ orbitals. Second, tunneling in the $z$-direction is
further suppressed by a linear tilt of the energy per lattice site, {making}
the system dynamically 2D. The $p$-orbital fermions here can be described by the
following multi-orbital model in the tight binding regime,

\begin{figure}[t]
\includegraphics[width=9cm]{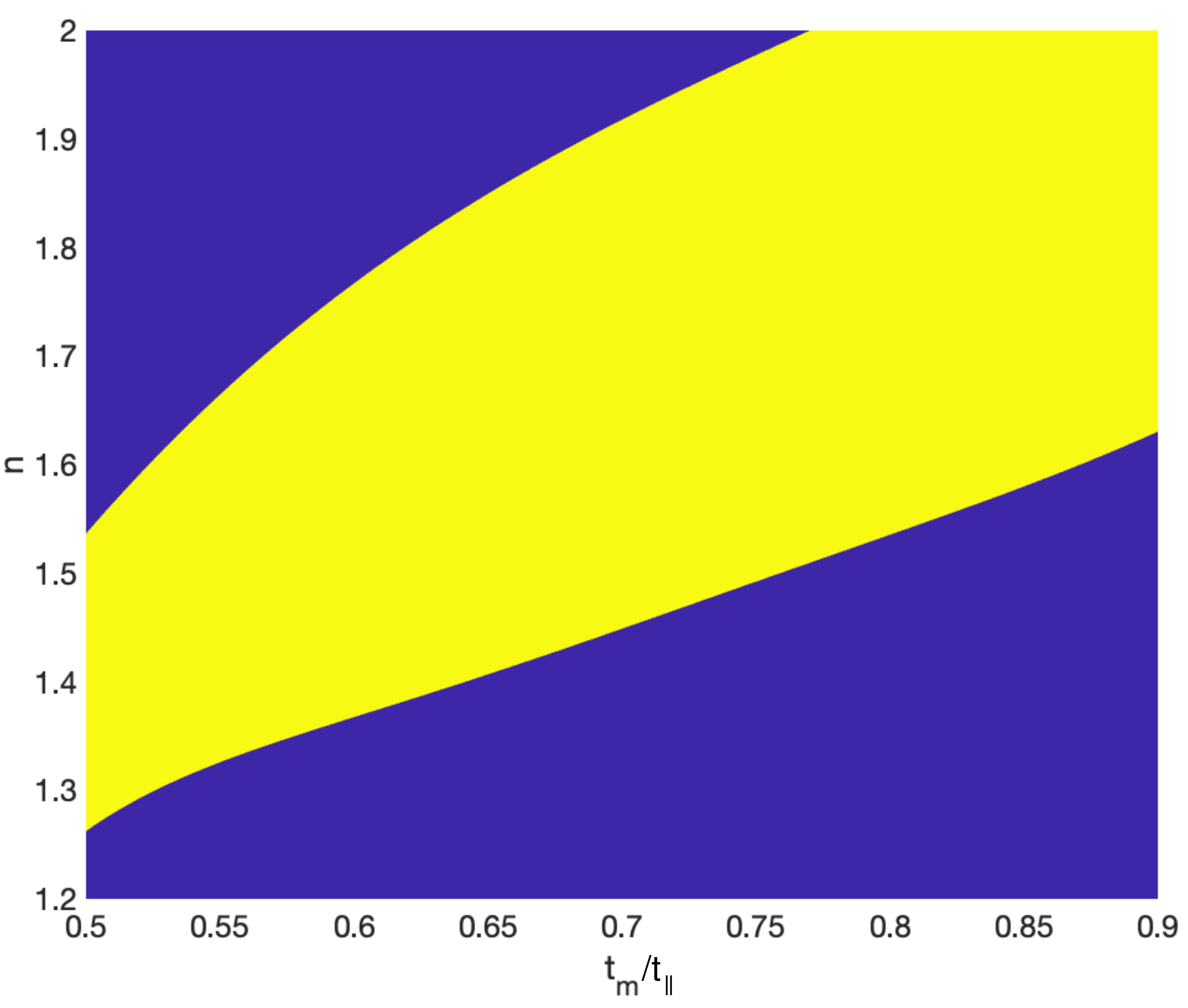}
\caption{Topological phase diagram as a function of orbital mixing strength $t_m$
for different filling $n$ of fermionic atoms in optical lattices.
The topological non-trivial (yellow area) and trivial (purple area) superfluids are separated and the phase boundary corresponds with the closing of the bulk gap. Interestingly,
the topological superfluid proposed here {possesses} high Chern number, i.e., $C=2$. Other parameters are chosen as $|W|=0.5t_{\parallel}$, $U=3W$, $t_z=t_{\perp}=0.2t_{\parallel}$ and $\Omega_z=0.5t_{\parallel}$.}
\label{fig:TZeroPhaseDiagDensity}
\end{figure}

\begin{widetext}
\begin{eqnarray}
\mathbf{H}_{0} & = & t_{\parallel}\sum_{\mathbf{r_i}}C_{p_{x}}^{\dagger}\left(\mathbf{r_i}\right)C_{p_{x}}\left(\mathbf{r_i}+\vec{e}_{x}\right)-t_{\perp}\sum_{\mathbf{r_i}}C_{p_{x}}^{\dagger}\left(\mathbf{r_i}\right)C_{p_{x}}\left(\mathbf{r_i}+\vec{e}_{y}\right)+t_{\parallel}\sum_{\mathbf{r_i}}C_{p_{y}}^{\dagger}\left(\mathbf{r_i}\right)C_{p_{y}}\left(\mathbf{r_i}+\vec{e}_{y}\right)-\nonumber \\
 &  & t_{\perp}\sum_{\mathbf{r_i}}C_{p_{y}}^{\dagger}\left(\mathbf{r_i}\right)C_{p_{y}}\left(\mathbf{r_i}+\vec{e}_{x}\right)-t_{z}\sum_{\mathbf{r_i}}\left[C_{p_{z}}^{\dagger}\left(\mathbf{r_i}\right)C_{p_{z}}\left(\mathbf{r_i}+\vec{e}_{x}\right)+C_{p_{z}}^{\dagger}\left(\mathbf{r_i}\right)C_{p_{z}}\left(\mathbf{r_i}+\vec{e}_{y}\right)\right]+\nonumber \\
 &  & t_{m}\sum_{\mathbf{r_i}}\left[C_{p_{x}}^{\dagger}\left(\mathbf{r_i}+\vec{e}_{x}\right)C_{p_{z}}\left(\mathbf{r_i}\right)-C_{p_{x}}^{\dagger}\left(\mathbf{r_i}\right)C_{p_{z}}\left(\mathbf{r_i}+\vec{e}_{x}\right)\right]+\nonumber \\
 &  & t_{m}\sum_{\mathbf{r_i}}\left[C_{p_{y}}^{\dagger}\left(\mathbf{r_i}+\vec{e}_{y}\right)C_{p_{z}}\left(\mathbf{r_i}\right)-C_{p_{y}}^{\dagger}\left(\mathbf{r_i}\right)C_{p_{z}}\left(\mathbf{r_i}+\vec{e}_{y}\right)\right]+h.c.,
\label{eq:LatticeHamiltonian}
\end{eqnarray}
\end{widetext}
where $C_{\nu}^{\dagger}\left(\mathbf{r_i}\right)$ and $C_{\nu}\left(\mathbf{r_i}\right)$
with $\nu$ = $p_{x}$, $p_{y}$, $p_{z}$, are fermionic
creation and annihilation operators for the particle in the $\nu$
orbital at lattice site $\mathbf{r_i}$. $t_{\parallel}$ is the longitudinal hopping
and $t_{\perp}$ is the corresponding transverse hopping.
The relative signs of the hopping
amplitudes are fixed by the parity of $p_{x}$ and $p_{y}$ orbitals. $t_{z}$
describes the hopping of the $p_{z}$ fermions in the $xy$-plane. The key ingredient
in our model is the hybridization between the $p_{x}$ and $p_{z}$, $p_{y}$ and
$p_{z}$ orbitals, characterized by $t_{m}$ in Eq.~\eqref{eq:LatticeHamiltonian}.
Such hybridization between degenerate orbitals arises from the asymmetric
shape of the $p_{z}$ orbital wavefunction induced by the inversion symmetry
breaking along the $z$-direction, {which is highly tunable through varying the magnetic field gradient as shown in Fig. S1 of the Supplemental Material (SM).} In the momentum space, the single-particle Hamiltonian
Eq.~\eqref{eq:LatticeHamiltonian} can be rewritten as $$\mathbf{H}_{0}\left( \mathbf{k}
\right) =\left( C_{p_{x}}^{\dagger }\left( \mathbf{k}\right)
C_{p_{y}}^{\dagger }\left( \mathbf{k}\right) C_{p_{z}}^{\dagger }\left(
\mathbf{k}\right) \right) \mathbf{\mathscr{H}}\left( \mathbf{k}\right) \left(
\begin{array}{c}
C_{p_{x}}\left( \mathbf{k}\right)  \\
C_{p_{y}}\left( \mathbf{k}\right)  \\
C_{p_{z}}\left( \mathbf{k}\right)
\end{array}%
\right), $$ where
\begin{widetext}
\begin{equation}
\mathbf{\mathscr{H}}\left(\mathbf{k}\right)= \left(\begin{array}{ccc}
2t_{\parallel}\cos\left(k_{x}a\right)-2t_{\perp}\cos\left(k_{y}a\right) & 0 & 2it_{m}\sin\left(k_{x}a\right)\\
0 & 2t_{\parallel}\cos\left(k_{y}a\right)-2t_{\perp}\cos\left(k_{x}a\right)& 2it_{m}\sin\left(k_{y}a\right)\\
-2it_{m}\sin\left(k_{x}a\right) & -2it_{m}\sin\left(k_{y}a\right) & -2t_{z}\left(\cos\left(k_{x}a\right)+\cos\left(k_{y}a\right)\right)
\end{array}\right),
\label{eq:MomentumSpace3x3}
\end{equation}
\end{widetext}
with $\mathbf{k}=\left(k_{x},k_{y}\right)$ taking values in the first
Brillouin zone and $a\equiv a_x=a_y$. $C_{\nu}^{\dagger}\left(\mathbf{k}\right)$ and $C_{\nu}\left(\mathbf{k}\right)$
represent the $\nu$-orbital fermionic creation and annihilation operators in  momentum space, respectively. The eigenvalues of the $3 \times 3$ matrix in Eq.~\eqref{eq:MomentumSpace3x3} give the {single-particle}
band structure. As shown in Fig.~\ref{fig:QuiverPlot}(a), we find that the second
and the third band cross at $\Gamma(k_{x}=0,k_{y}=0)$ point (the center of the Brillouin zone). {Zooming into the vicinity of $\Gamma$ shows that it develops a parabolic touching, which is qualitatively distinct from the linear Dirac or Weyl points.} Surprisingly, such a band degeneracy point
in our model behaves like a topological defect, i.e., a vortex with winding number $2$, and thus an unexpected topological semimetal {is developed}. To show the topological nature of such a semimetal, we derive an effective low-energy theory around the band degeneracy point at $\Gamma$ (see details in SM), captured by the following effective Hamiltonian
\begin{equation}
\mathbf{H}_{eff}=c_{0}\left(k_{x}^{2}+k_{y}^{2}\right)\mathbb{I}_{2\times2}+c_{1}k_{x}k_{y}%
\sigma_{x}+c_{3}\left(k_{x}^{2}-k_{y}^{2}\right)\sigma_{z},
\label{eq:Effective2x2Hamiltonian}
\end{equation}
where $c_{0}={a^2}[-t_{\parallel}+t_{\perp}+{2t_{m}^{2}}/({t_{\parallel}-t_{\perp}+2t_{z}})]/2$,
$c_{1}={2t_{m}^{2}{a^2}}/({t_{\parallel}-t_{\perp}+2t_{z}})$ and $c_{3}={a^2}[-t_{\parallel}-t_{\perp}+{2t_{m}^{2}}/({t_{\parallel}-t_{\perp}+2t_{z}})]/2$.
{($\mathbb{I}_{2\times2}$, $\sigma_{x}$, $\sigma_{z}$) are the unit and Pauli matrices.} To visualize the topological nontriviality of the quadratic
touching point here, a 2D vector field $\mathbf{h}\left(\mathbf{k}\right)$ can be defined from the coefficients of the two Pauli matrices in
$\mathbf{H}_{eff}$ as
\begin{equation}
\mathbf{h}\left(\mathbf{k}\right)=\left(%
\begin{array}{c}
c_{1}k_{x}k_{y},
c_{3}\left(k_{x}^{2}-k_{y}^{2}\right)%
\end{array}%
\right).  \label{eq:TwoVectorHVortex}
\end{equation}
As shown in Fig.~\ref{fig:QuiverPlot}(b), the vector field $\mathbf{h}\left(\mathbf{k}\right)$ forms a vortex
structure in the momentum space. At the vortex core, the length of the vector vanishes, indicating
that the band gap vanishes at the band degeneracy point. Therefore, such a band-touching point forms a topological defect. Its topological nature is also confirmed
from the calculation of the winding number (see details in SM) and
it turns out that the winding number is $2$, {which is distinct} from the linear Dirac point with winding number $1$. In fact, the band degeneracy point here is protected by the lattice rotation $C_4$ and reflection symmetries, where the
reflection in the horizontal and vertical directions are associated
with the transformations of fermionic operators as
$\mathcal{R}_{\parallel}\equiv\{C_{p_{x}}(\mathbf{r_i})\rightarrow -C_{p_{x}}(-\mathbf{r_i}_{x},\mathbf{r_i}_{y}), C_{p_{y(z)}}(\mathbf{r_i})\rightarrow
C_{p_{y(z)}}(-\mathbf{r_i}_{x},\mathbf{r_i}_{y})\}$ and $\mathcal{R}_{\perp} \equiv \{
C_{p_{x(z)}}(\mathbf{r_i})\rightarrow C_{p_{x(z)}}(\mathbf{r_i}_{x},-\mathbf{r_i}_{y}), C_{p_{y}}(\mathbf{r_i})\rightarrow
-C_{p_{y}}(\mathbf{r_i}_{x},-\mathbf{r_i}_{y})\}$, respectively. {{Those} associated with the $\pi/2$-lattice rotation can be expressed as $C_{p_{x(z)}}(\mathbf{r_i})\rightarrow C_{p_{y(z)}}(-\mathbf{r_i}_{y},\mathbf{r_i}_{x}), C_{p_{y}}(\mathbf{r_i})\rightarrow
-C_{p_{x}}(-\mathbf{r_i}_{y},\mathbf{r_i}_{x})$. Here we would like to emphasize that {the essential ingredient of} our proposed topological semimetal arises from the non-trivial hybridization between different orbitals. The key idea in our scheme is to utilize the unique and intrinsic anisotropic spatial nature of higher oribtals. Through simply employing a magnetic field gradient to break the spatial inversion symmetry, the topologically non-trivial mixing between orbitals is {thus} induced, resulting in a substantially novel way of producing the topological semimetal in the ultracold atom based system.

\textit{$S$-wave interaction induced topological superfluid with high Chern number$\raisebox{0.01mm}{---}
$} {In the previous section, it is shown that the non-trivial orbital mixing produces a topological band featured with the quadratic topological
defect. Past studies show that it is unstable to the repulsive interaction,
resulting in various topological phases, such as topological insulators and
nematic phases~\cite{kaisun_PhysRevB,2012_Kaisun_naturephy,Oganesyan_PhysRevB,Kivelson_RevModPhys,Kaisun_PhysRevLett}. However, when considering the instability driven by the attraction, the singlet superconducting pairing is absent ~\cite{Kaisun_PhysRevLett,2018_wang_CPB}. Here we show
that when adding a new ingredient, i.e., on-site rotation, a new type of topological
superfluid with high Chern number can be created via the $s$-wave attraction induced singlet pairing. It thus opens up a new thrust towards exploring topological superfluids directly from an $s$-wave interaction, yet without requiring Raman-induced spin-orbit coupling nor other artificial gauge field. To illustrate this, let us consider loading spin-1/2 attractive fermionic atoms into the lattice system described above. With the addition of on-site rotation, the interacting model can be expressed as}

\begin{equation}
\textstyle
\mathbf{H}=\mathbf{H}_{0,\sigma}+\mathbf{H}_{int}+\mathbf{H}_{L}.
\label{Haminteraction}
\end{equation}
The interaction part $\mathbf{H}_{int}$ is
\begin{widetext}
\begin{eqnarray}
\mathbf{H}_{int}&=&\sum\limits_\mathbf{r_i}\big\{U\sum_{\nu}n\left(\mathbf{r_i}\right)_{\nu,\uparrow}n\left(\mathbf{r_i}\right)_{\nu,\downarrow}
+W\sum\limits_{\nu\neq \rho}\big[n\left(\mathbf{r_i}\right)_{\nu,\uparrow}
n\left(\mathbf{r_i}\right)_{\rho,\downarrow}
+C^\dag_{\nu,\uparrow}\left(\mathbf{r_i}\right)C^\dag_{\rho,\downarrow}\left(\mathbf{r_i}\right)C_{\nu,\downarrow}\left(\mathbf{r_i}\right)C_{\rho,\uparrow}\left(\mathbf{r_i}\right)\nonumber\\
&+&C^\dag_{\nu,\uparrow}\left(\mathbf{r_i}\right)C^\dag_{\nu,\downarrow}\left(\mathbf{r_i}\right)C_{\rho,\downarrow}\left(\mathbf{r_i}\right)C_{\rho,\uparrow}\left(\mathbf{r_i}\right)\big]\big\},
\label{eq:InteractionTerm}
\end{eqnarray}
\end{widetext}
where the onsite particle number operator is defined as $n_{\nu,\sigma}\left(\mathbf{r_i}\right)\equiv C^\dag_{\nu,\sigma}\left(\mathbf{r_i}\right)C_{\nu,\sigma}\left(\mathbf{r_i}\right)$
and $\nu,\rho =p_x, p_y, p_z$.
The {intra-orbital and inter-orbital} interacting strength are captured by $U$ and $W$, respectively.
In general, $U\neq W$ which results from the anisotropic shape of $p$-orbitals. These onsite attraction strengths are determined by the effective $s$-wave scattering lengths, which can be tuned by means of {the Feshbach resonance and lattice depth}. $H_{L}=i\Omega_z\sum_{{\mathbf {r_i}},\sigma}[C_{p_x,\sigma}^{\dagger
}(\mathbf{r_i})C_{p_y,\sigma}(\mathbf{r_i})-C_{p_y,\sigma}^{\dagger
}(\mathbf{r_i})C_{p_x,\sigma}(\mathbf{r_i})]$ is induced by the
on-site rotation. Such an on-site rotation experiment has been achieved in a triangular
optical lattice~\cite{GemelkeThesis,GemelkeChuPaper} and the techniques are
expected to be applicable to the model considered here. Through utilizing electro-optic phase modulators of the laser beams forming the lattice, we propose a method to realize the on-site rotation in our setup (see SM for details).

\begin{figure}[t]
\includegraphics[width=9cm]{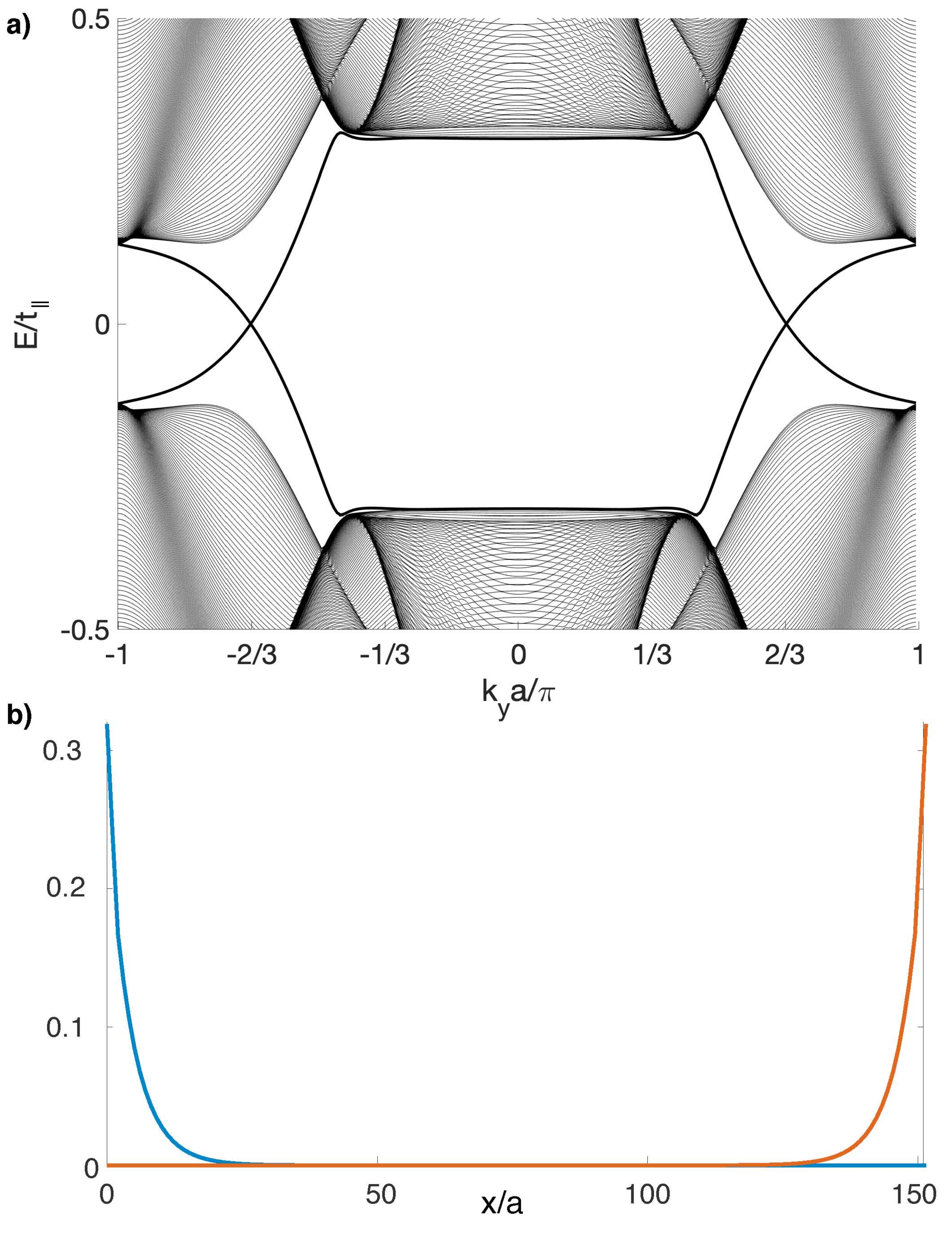}
\caption{(a) Energy spectrum of the Hamiltonian in Eq.~\eqref{Haminteraction}
with open (periodic) boundary condition in the x (y) direction. There are two
pairs of chiral modes located at the two outer edges, respectively.
(b) The amplitudes of wavefunctions for the two chiral edge modes
with $k_ya = 2\pi/3$ in (a).
Here we choose $t_m=0.8t_{\parallel}$ and $n=1.6$. Other parameters are chosen as the same in Fig.~\ref{fig:TZeroPhaseDiagDensity}.}
\label{fig:EdgeStates}
\end{figure}

\begin{figure}[t]
\includegraphics[width=9.5cm]{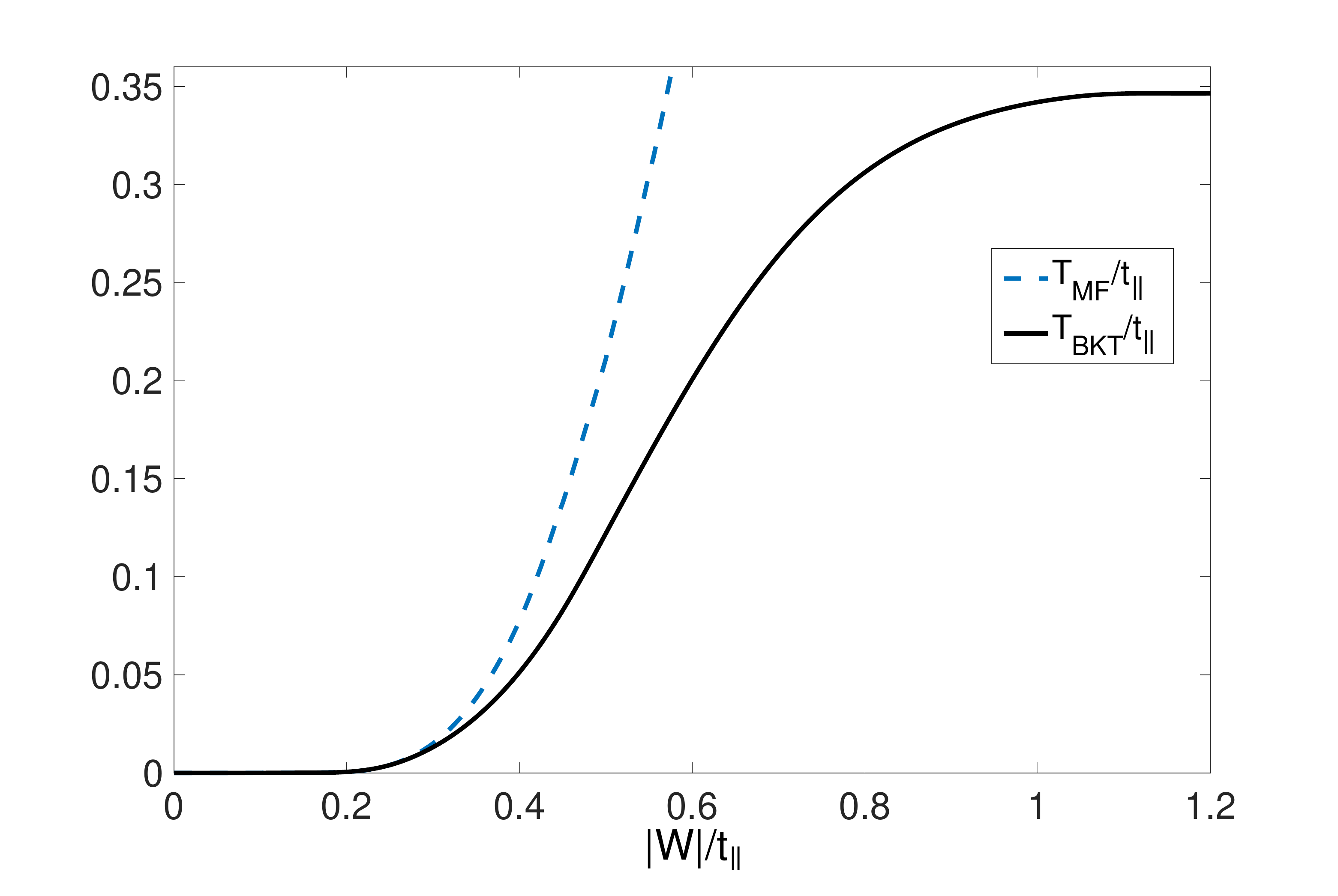}
\caption{BKT transition temperature $T_{BKT}$ as a function of interaction
strength. For comparison, the mean-field transition temperature $T_{MF}$ is
also shown here. We choose $t_m=0.8t_{\parallel}$ and $n=1.6$. Other parameters are the same as in Fig.~\ref{fig:EdgeStates}.}
\label{fig:TBKT}
\end{figure}
Due to the attractive interactions, the fermions tend to pair with each
other and form a superfluid state at low temperatures. To study this
superfluid state, we construct a path-integral formalism. The details
are given in SM. The auxiliary complex bosonic fields
$\Delta^{(j)}_{\nu\rho}(\mathbf{r_i},\tau)$ and $\overline{\Delta}^{(j)}_{\nu\rho}(\mathbf{r_i},\tau)$ with $\nu,\rho=p_{x},p_{y},p_{z}$ are introduced to obtain a bosonic effective action by Hubbard-Stratonovich transformation. Under the saddle-point approximation,
the pairing order parameters $\Delta_{\nu\rho}$ can be determined (see details
in SM).  Surprisingly, we find that by simply tuning the average
filling $n$ of fermions in such a lattice system, the superfluids with {distinct} topological properties can be achieved. The topological nature of
superfluidity here can be characterized by its non-trivial topological invariant. At mean-field level, the system can be described by the Bogoliubov-de Gennes (BdG) Hamiltonian ({see SM for details}). Since such a BdG Hamiltonian satisfies the particle-hole symmetry, i.e., $\Xi H_{BdG}(\mathbf
{k})\Xi^{-1}=-H_{BdG} ^{*} (\mathbf {-k})$, with
$\Xi\equiv \begin{pmatrix} {0} & {\mathbb{I}_{3\times3}}\\ {\mathbb{I}_{3\times3}} & {0}
\end{pmatrix}\otimes\sigma_{y}$, the superfluid state predicted here belongs
to the $D$ symmetry class according to the general classification
scheme of topological superconductors~\cite{2008_Schnyder_PhysRevB}.
Therefore, its topology can be studied by evaluating the
Chern number. {Fig.~\ref{fig:TZeroPhaseDiagDensity} shows
the topological phase diagram of the proposed superfluids.}
There are two different topological regions in the phase diagram,
which are characterized by different Chern numbers. Intriguingly,
a new topological orbital-hybridized superfluid with high
Chern number, i.e., $C=2$, is unveiled. {For a fixed interaction
strength, the system undergoes unusual type of topological phase transition from a
topological trivial superfluid (SF) to the $C = 2$ topological non-trivial
superfluid (tSF) when varying the filling of fermions.}
As shown in Fig.~\ref{fig:TZeroPhaseDiagDensity}, the tSF phase region is
quite sizable and would make the experimental realization easier. To further demonstrate the topological nature of tSF
phase, we shall show that the chiral edge modes are supported
in this state. To see this, we consider a cylinder
geometry of the system, where the open (periodic) boundary
condition is chosen in the x(y) direction, respectively. The
energy spectrum in Fig.~\ref{fig:EdgeStates} is labeled by momentum $k_y$.
As shown in Fig.~\ref{fig:EdgeStates}(a), all the bulk modes are gapped and there
are two pairs of chiral edge states
located at the two outer edges of the system, respectively,
which are confirmed by the well-localized wave functions
as shown in Fig.~\ref{fig:EdgeStates}(b). It satisfies the so-called bulk-edge
correspondence, since the Chern number of tSF state
is $2$. Since the number of chiral edge modes determines the quantized conductance in the system~\cite{RevModPhys1,RevModPhys2}, the high Chern numbers are preferable in achieving more efficient edge channel transport. Therefore, our scheme would shed light on the new possibilities for edge-mode engineering through the fabrication of topological phases {in both electronic solid state and atomic gas matter
~\cite{Joel_PhysRevLett,Bernevig_NC,Joel_PhysRevB,Gong_PhysRevB,Ye_PhysRevB}.}

In the superfluid region, as the temperature increases, the
system eventually undergoes a Berezinskii-Kosterlitz-Thouless
(BKT) transition from the superfluid to normal state.
At the BKT critical temperature~\cite{1972_Berezinskii,1973_KosterlitzThouless},
the vortex-antivortex pairs disassociate
and it costs zero free energy to generate a single unbound vortex.
The finite temperature phase diagram of our proposed system is obtained
in Fig.~\ref{fig:TBKT} (see details in SM). In the current experiments, such as $^6$Li or
$^{40}$K~\cite{2004_Jin_PRL,2004_Martin_PRL,2004_Tomas_PRL,
2004_Grimm_PRL,2004_Salomon_PRL}, taking advantage of the experimental realization of Feshbach
resonance, the interaction is highly tunable. The BKT transition
temperature can be estimated to reach around $40$nK, making it promising to
obtain the proposed tSF phase in experiments.

\textit{Conclusions $\raisebox{0.01mm}{---}$} We have demonstrated a symmetry-based systematic method of engineering the non-trivial orbital hybridization in optical lattices. Various unexpected orbital-hybridized topological phases, including a topological semimetal with quadratic band touching, as well
as a topological superfluid with high Chern number, are predicted through our proposed scheme. Moreover, this mechanism can be easily achieved in current experiments
by utilizing our protocol of manipulating inversion symmetry} in the optical lattice system, potentially circumventing the challenges in Raman-induced spin-orbit coupling scheme.
The present approach thus complements with a new window to investigate topological phases in cold gases.

\textit{Acknowledgments} This work is supported by the National Key Research and Development Program of China (2018YFA0307600), NSFC (Grant No. 12074305, 11774282, 11950410491), Cyrus Tang Foundation Young Scholar Program and the Fundamental Research Funds for the Central Universities (M. A., S. L. and B. L.), and by AFOSR
Grant No.~FA9550-16-1-0006, MURI-ARO Grant No.~W911NF-17-1-0323 through UC Santa
Barbara, and Shanghai Municipal Science and Technology Major Project (Grant
No. 2019SHZDZX01) (W.V. L.).

\bibliographystyle{apsrev}
\bibliography{HighChernNumberBIB}

\onecolumngrid

\renewcommand{\thesection}{S-\arabic{section}}
\setcounter{section}{0}  
\renewcommand{\theequation}{S\arabic{equation}}
\setcounter{equation}{0}  
\renewcommand{\thefigure}{S\arabic{figure}}
\setcounter{figure}{0}  

\indent

\begin{center}\large
\textbf{Supplementary Material:\\Topological semimetal and superfluid of $s$-wave interacting fermionic atoms in an orbital optical lattice}
\end{center}
\section{orbital-hybridized topological semimetal\label{sec:topsemimetal}}

\begin{figure}[t]
\includegraphics[width=12cm]{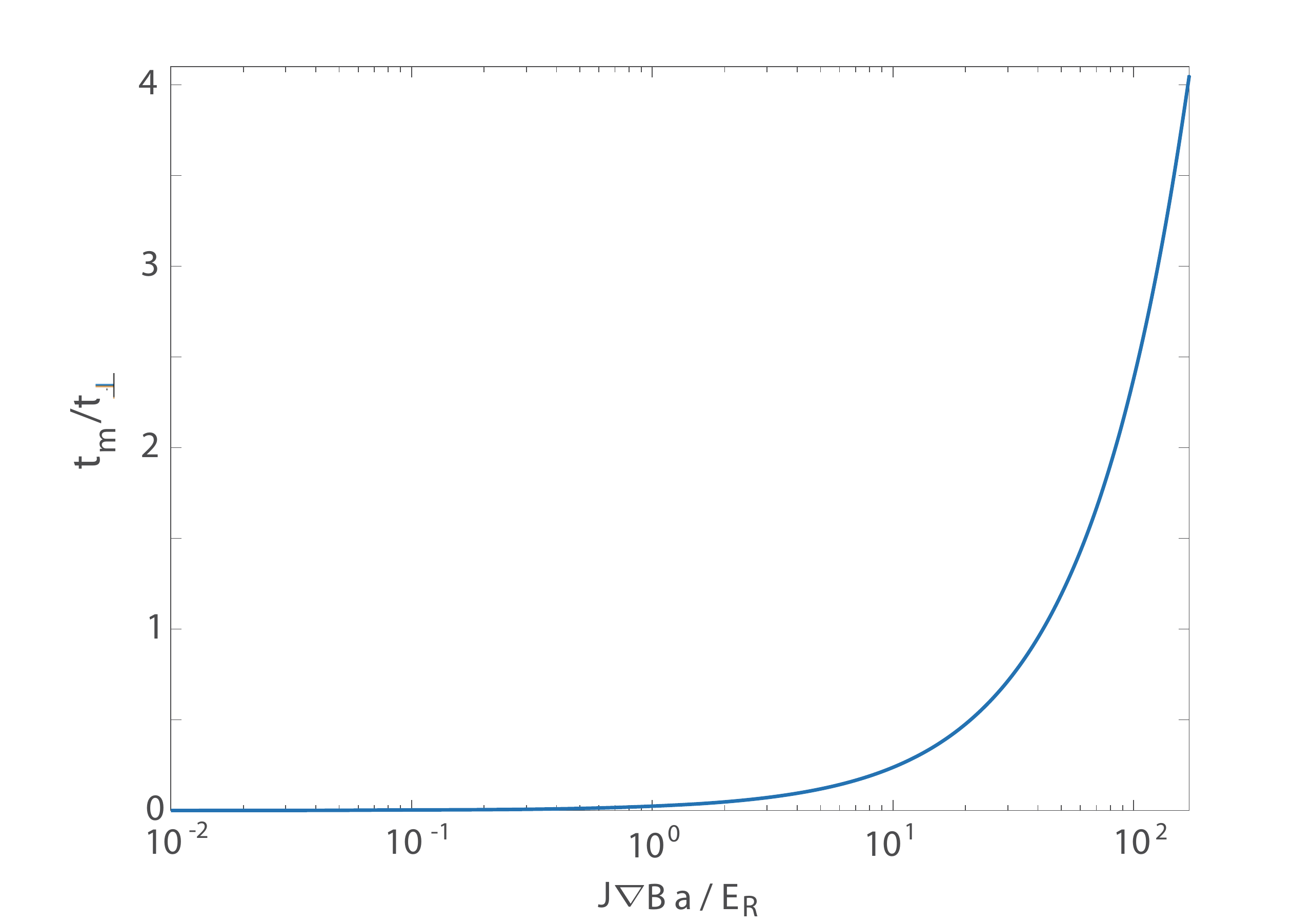}
\caption{The orbital hybridization $t_{m}/t_{\perp}$ as a function of the
magnetic field gradient. Here $E_R$ is the recoil energy defined as $E_R=\frac{\hbar^2 k^2_{L}}{2m}$. }
\label{fig:Orbitalmix}
\end{figure}

In this section, we shall provide a detailed {derivation of} the effective Hamiltonian in Eq. (3). The method of Feshbach projectors \cite{FeshbachProjectors} is used here to reduce the model in Eq. (2) to an effective low-energy Hamiltonian around the band touching point. The projection operators $\mathbf{P}$ and $\mathbf{Q}=1-\mathbf{P}$ are introduced that project into the subspaces spanned by the two touching bands at the $\Gamma$ point and the remaining band, respectively. The eigenvalue problem associated with the model Hamiltonian $\mathbf{H_0}$ can be written as $\mathbf{H_0}|\Psi \rangle =E|\Psi \rangle$,
where $|\Psi \rangle$ is the eigenstate with eigenenergy $E$. The effective Hamiltonian $\mathbf{H}_{eff}$ which describes the low-energy physics around the band touching point $\Gamma$ can be obtained through
\begin{equation}
\left( \mathbf{PH_{0}P+PH_{0}Q}\frac{1}{E-\mathbf{QH_{0}Q}}\mathbf{QH_{0}P}\right) \mathbf{P}|\Psi \rangle \equiv \mathbf{H}_{eff}\mathbf{P}|\Psi \rangle=E\mathbf{P}|\Psi \rangle .
\label{eq:ProjectedEquation}
\end{equation}%
By computing the relevant matrix elements of Eq. ~\eqref{eq:ProjectedEquation} in the projected Hilbert space, the effective Hamiltonian $\mathbf{H}_{eff}$ can be further expressed as  \begin{eqnarray}
\mathbf{H}_{eff} &=&\left(
\begin{array}{cc}
2t_{\parallel}\cos \left( k_{x}a\right) -2t_{\perp}\cos \left( k_{y}a\right)  & 0 \\
0 & 2t_{\parallel}\cos \left( k_{y}a\right) -2t_{\perp}\cos \left( k_{x}a\right)
\end{array}%
\right)   \notag \\
&&+\frac{2t_{m}^{2}}{t_{\parallel}-t_{\perp}+2t_{z}\left(\cos\left(k_x a\right)+\cos\left(k_y a\right)\right)}\left(
\begin{array}{cc}
\sin ^{2}\left( k_{x}a\right)  & \sin \left( k_{x}a\right) \sin \left(
k_{y}a\right)  \\
\sin \left( k_{x}a\right) \sin \left( k_{y}a\right)  & \sin ^{2}\left(
k_{y}a\right)
\end{array}%
\right) .  \label{eq:Heff}
\end{eqnarray}%
Expanding Eq. ~(S2) up to the second order in momentum near the $\Gamma$ point, $\mathbf{H}_{eff}$
can be further approximated as
\begin{equation}
\mathbf{H}_{eff}\left( k_{x},k_{y}\right) \approx  c_{0}\left( k_{x}^{2}+k_{y}^{2}\right) \mathbb{I}_{2\times2}+c_{1}k_{x}k_{y}\sigma
_{x}+c_{3}\left( k_{x}^{2}-k_{y}^{2}\right) \sigma _{z},
\label{eq:PauliNotation}
\end{equation}%
where
$c_{0}=\frac{a^2}{2}\left( -t_{\parallel}+t_{\perp}+\frac{2t_{m}^{2}}{t_{\parallel}-t_{\perp}+2t_{z}}%
\right),
c_{1}=\frac{2t_{m}^{2} a^2}{t_{\parallel}-t_{\perp}+2t_{z}}$,  and $c_{3}=\frac{a^2}{2}\left( -t_{\parallel}-t_{\perp}+\frac{2t_{m}^{2}}{t_{\parallel}-t_{\perp}+2t_{z}}%
\right)$. $\sigma _{x}$, $\sigma _{z}$ are the Pauli matrices and
$\mathbb{I}_{2\times2}$ is the unit matrix. Eq. ~\eqref{eq:PauliNotation} can also be parameterized by a
vector in the $\left( k_{x},k_{y}\right)$-plane as
\begin{equation}
\mathbf{h}\left(\mathbf{k}\right)=\left(%
\begin{array}{c}
c_{1}k_{x}k_{y},
c_{3}\left(k_{x}^{2}-k_{y}^{2}\right)%
\end{array}%
\right).  \label{eq:TwoVectorHVortex}
\end{equation}
The vector $\mathbf{h}\left( k_{x},k_{y}\right)$ defined here has a vortex
structure with the winding number
$W=\ointop_{C}\frac{\mathrm{d}\mathbf{k}}{2\pi }\,\left[ \frac{h_{x}}{\left\vert
\mathbf{h}\right\vert }\nabla \left( \frac{h_{y}}{\left\vert \mathbf{h}%
\right\vert }\right) -\frac{h_{y}}{\left\vert \mathbf{h}\right\vert }\nabla
\left( \frac{h_{x}}{\left\vert \mathbf{h}\right\vert }\right) \right]$
being equal to $2$.

\section{Local rotation} \label{sec:localrotation}
Through putting electro-optic modulators on the laser beams
forming the lattice~\cite{GemelkeThesis,GemelkeChuPaper},
the lattice potential can be expressed as

\begin{eqnarray}
V(x,y,z)&=& -\frac{V}{2}\left[\cos^{2}\left(k_{L}x+\phi_{x}\left(t\right)\right)+\cos^{2}\left(k_{L}y+\phi_{y}\left(t\right)\right)\right]\nonumber \\
&-&\frac{V}{4}\left[\cos^{2}\left(k_{L}x+k_{L}y+\phi_{+}\left(t\right)\right)+\cos^{2}\left(k_{L}x-k_{L}y+\phi_{-}\left(t\right)\right)\right]-V_{z}\cos^{2}\left(k_{L_{z}}z\right),
\label{eq:ModulatedPotentialAppendix}
\end{eqnarray}
where $V \equiv V_x=V_y$, $k_{L}\equiv k_{L_x}=k_{L_y}$ and the electro-optic phase modulators $\phi_{x}\left(t\right)=\Delta\phi\cos(\Omega_{z}t)\cos(\omega_{RF}t)$,
$\phi_{y}(t)=\Delta\phi\cos(\Omega_{z}t+\pi/2)\cos(\omega_{RF}t)$,
$\phi_{+}(t)=\Delta\phi\cos(\Omega_{z}t+\pi/4)\cos(\omega_{RF}t)$,
$\phi_{-}(t)=\Delta\phi\cos(\Omega_{z}t-\pi/4)\cos(\omega_{RF}t)$ with the slow
precession frequency ${\Omega_z}$, the amplitude of oscillation
$\Delta\phi$ and the fast rotation frequency $\omega_{RF}$ at radio
frequency.
It results in a periodical overall translation of the lattice
at a radio-frequency $\omega_{RF}$. Atoms do not follow the fast
oscillation at radio frequency $\omega_{RF}$ and only feel a time
averaged potential. The local potential near each site minimum in
the rotating frame with frequency ${\Omega_z}$ can be approximately
(dropping a constant) expressed as
\begin{equation}
V\left( x',y',z\right) \approx -\frac{3V}{2}\left(1-\frac{\Delta\phi^{2}}{4}\right)+\frac{m\omega'^{2}r'^{2}}{2}\left(1+2\epsilon\cos\left(2\phi'\right)\right)+V_{z}k^2_{L_z}z^2,
\label{eq:RotatedPotential}
\end{equation}%
where $\frac{m\omega'^{2}}{2}=Vk_{L}^{2}\left(1-\frac{\Delta\phi^{2}}{2}\right)$, $\epsilon=-\frac{\Delta\phi^{2}}{8\left(1-\Delta\phi^{2}/2\right)}$, $r'^{2}=x'^{2}+y'^{2}$ and $\phi'$
is the polar angle of ${\mathbf r'}$. The reference frame with rotating axis along $z$-direction and angular velocity $\Omega _{z}$ with respect to the original frame can be captured by the following transformation
$\left(
\begin{array}{c}
x^{\prime } \\
y^{\prime }%
\end{array}%
\right) =\left(
\begin{array}{cc}
\cos (\Omega _{z}t) & -\sin (\Omega _{z}t) \\
\sin (\Omega _{z}t) & \cos (\Omega _{z}t)%
\end{array}%
\right) \left(
\begin{array}{c}
x \\
y%
\end{array}%
\right)$. The slight deformation of the
optical potential processes around each site center, i.e.,
the fourth term in Eq. (S6), which can be regarded as an on-site rotation~\cite{GemelkeThesis,GemelkeChuPaper}.

{\section{path integral formalism}}
By introducing Grassmann fields $\bar{C}_{\mu(\nu),\sigma}\left(\mathbf{r_{i}},\tau\right)$,
$C_{\mu(\nu),\sigma}\left(\mathbf{r_{i}},\tau\right)$ with $\mu(\nu)=p_{x},p_{y},p_{z}$ and $\sigma=\uparrow,\downarrow$, which represent fermionic fields, the partition function of the system can be expressed as (the units are chosen as $%
\hbar =k_{B}=1$)
\begin{equation}
\mathcal{Z}=\int\mathcal{D}\bar{C}\mathcal{D}C\,\exp\left(-S\left[\bar{C},C\right]\right),\label{eq:PartitionFunction}
\end{equation}
with the action $S$
\begin{equation}
S=S_{0}\left[\bar{C},C\right]+S_{int}\left[\bar{C},C\right],\label{eq:Action}
\end{equation}
where
\begin{equation*}
S_{0}\left[\bar{C},C\right]=\intop_{0}^{\beta} \mathrm{d}\tau\sum_{\mu\nu\mathbf{r_{i}r'_{i}}\sigma}\bar{C}_{\mu,\sigma}\left(\mathbf{r_{i}},\tau\right)\left(\delta_{\mu\nu}\delta_{\mathbf{r_{i}},\mathbf{r'_{i}}}\partial_{\tau}+{\mathscr{H}}_{\mu\nu}\right)C_{\nu,\sigma}\left(\mathbf{r'_{i}},\tau\right),
\end{equation*}
\begin{equation} \label{eq:ActionExpanded}
\begin{aligned}
S_{int}\left[\bar{C},C\right]=&W\intop_{0}^{\beta} \mathrm{d}\tau\sum_{\mathbf{r_{i}}}\left\{ 3\sum_{\mu}\bar{C}_{\mu,\uparrow}\left(\mathbf{r_{i}},\tau\right)\bar{C}_{\mu,\downarrow}\left(\mathbf{r_{i}},\tau\right)C_{\mu,\downarrow}\left(\mathbf{r_{i}},\tau\right)C_{\mu,\uparrow}\left(\mathbf{r_{i}},\tau\right)\right. \\
&\left.+\sum_{\mu\neq\nu}\bar{C}_{\mu,\uparrow}\left(\mathbf{r_{i}},\tau\right)\bar{C}_{\mu,\downarrow}\left(\mathbf{r_{i}},\tau\right)C_{\nu,\downarrow}\left(\mathbf{r_{i}},\tau\right)C_{\nu,\uparrow}\left(\mathbf{r_{i}},\tau\right)+\sum_{\mu\neq\nu}\bar{C}_{\mu,\uparrow}\left(\mathbf{r_{i}},\tau\right)\bar{C}_{\nu,\downarrow}\left(\mathbf{r_{i}},\tau\right)C_{\mu,\downarrow}\left(\mathbf{r_{i}},\tau\right)C_{\nu,\uparrow}\left(\mathbf{r_{i}},\tau\right)\right.\\
&\left.+\sum_{\mu\neq\nu}\bar{C}_{\mu,\uparrow}\left(\mathbf{r_{i}},\tau\right)\bar{C}_{\nu,\downarrow}\left(\mathbf{r_{i}},\tau\right)C_{\nu,\downarrow}\left(\mathbf{r_{i}},\tau\right)C_{\mu,\uparrow}\left(\mathbf{r_{i}},\tau\right)\right\},
\end{aligned}
\end{equation} with assuming $U=3W$.

The quartic fermionic interaction term in action $S$ can be decoupled by introducing Hubbard-Stratonovich fields
\begin{equation}
\bar{\Delta}_{\mu\nu}^{(j)}\left(\mathbf{r_{i}},\tau\right) \quad \text{and} \quad \Delta_{\mu\nu}^{(j)}\left(\mathbf{r_{i}},\tau\right), \quad \text{where} \quad j=\begin{cases}
1, & \mu=\nu\\
1,2 & \mu\neq\nu
\end{cases} \quad \text{and} \quad \mu,\nu=p_{x},p_{y},p_{z}.\label{eq:NewDefinition}
\end{equation}
Then the interaction part in the partition function can be rewritten as follows
\begin{equation*}
\begin{aligned}
\int\mathcal{D}\bar{C}\mathcal{D}C\exp\left(-W\intop_{0}^{\beta} \mathrm{d}\tau\sum_{\mathbf{r_{i}}}\big \{ 3\sum_{\mu}\bar{C}_{\mu,\uparrow}\left(\mathbf{r_{i}},\tau\right)\bar{C}_{\mu,\downarrow}\left(\mathbf{r_{i}},\tau\right)C_{\mu,\downarrow}\left(\mathbf{r_{i}},\tau\right)C_{\mu,\uparrow}\left(\mathbf{r_{i}},\tau\right)
\right.\\
+\sum_{\mu\neq\nu}\bar{C}_{\mu,\uparrow}\left(\mathbf{r_{i}},\tau\right)\bar{C}_{\mu,\downarrow}\left(\mathbf{r_{i}},\tau\right)C_{\nu,\downarrow}\left(\mathbf{r_{i}},\tau\right)C_{\nu,\uparrow}\left(\mathbf{r_{i}},\tau\right)
\big \}\Bigg )
\end{aligned}
\end{equation*}
\begin{equation}
\begin{aligned}
=\int\mathcal{D}\bar{C}\mathcal{D}C\prod_{\mu}\mathcal{D}\bar{\mathit{\Delta}}_{\mu\mu}^{(1)}\mathcal{D}\mathit{\Delta}_{\mu\mu}^{(1)}\int\exp&\left(\intop_{0}^{\beta} \mathrm{d}\tau\sum_{\mathbf{r_{i}}} \sum_{\mu\neq\nu}\big \{ \frac{3\bar{\mathit{\Delta}}_{\mu\mu}^{(1)}\left(\mathbf{r_{i}},\tau\right)\mathit{\Delta}_{\mu\mu}^{(1)}\left(\mathbf{r_{i}},\tau\right)}{W}+\frac{\bar{\mathit{\Delta}}_{\mu\mu}^{(1)}\left(\mathbf{r_{i}},\tau\right)\mathit{\Delta}_{\nu\nu}^{(1)}\left(\mathbf{r_{i}},\tau\right)}{W}\big \} \,\right.\\
&+\intop_{0}^{\beta} \mathrm{d}\tau\sum_{\mathbf{r_{i}}}\sum_{\mu\neq\nu}\big\{ \left[3\bar{\mathit{\Delta}}_{\mu\mu}^{(1)}\left(\mathbf{r_{i}},\tau\right)+\bar{\mathit{\Delta}}_{\nu\nu}^{(1)}\left(\mathbf{r_{i}},\tau\right)\right]C_{\mu,\downarrow}\left(\mathbf{r_{i}},\tau\right)C_{\mu,\uparrow}\left(\mathbf{r_{i}},\tau\right)\\
&+\left[3\mathit{\Delta}_{\mu\mu}^{(1)}\left(\mathbf{r_{i}},\tau\right)+\mathit{\Delta}_{\nu\nu}^{(1)}\left(\mathbf{r_{i}},\tau\right)\right]\bar{C}_{\mu,\uparrow}\left(\mathbf{r_{i}},\tau\right)\bar{C}_{\mu,\downarrow}\left(\mathbf{r_{i}},\tau\right)\big\} \Bigg ),
\end{aligned}
\end{equation}
and the term
\begin{equation}
\begin{aligned}
\int\mathcal{D}\bar{C}\mathcal{D}C\exp\bigg(-W\intop_{0}^{\beta} \mathrm{d}\tau\sum_{\mathbf{r_{i}}}\sum_{\mu\neq\nu}\big[&
\bar{C}_{\mu,\uparrow}\left(\mathbf{r_{i}},\tau\right)\bar{C}_{\nu,\downarrow}\left(\mathbf{r_{i}},\tau\right)C_{\mu,\downarrow}\left(\mathbf{r_{i}},\tau\right)C_{\nu,\uparrow}\left(\mathbf{r_{i}},\tau\right)\\
&+\bar{C}_{\mu,\uparrow}\left(\mathbf{r_{i}},\tau\right)\bar{C}_{\nu,\downarrow}\left(\mathbf{r_{i}},\tau\right)C_{\nu,\downarrow}\left(\mathbf{r_{i}},\tau\right)C_{\mu,\uparrow}\left(\mathbf{r_{i}},\tau\right)\big ]\bigg ),
\end{aligned}
\end{equation}
for instance, when considering $\mu,\nu=p_x, p_y$,
\begin{equation*}
\begin{aligned}
\int\mathcal{D}\bar{C}\mathcal{D}C\exp \bigg( -W&\left.\intop_{0}^{\beta} \mathrm{d}\tau \sum_{\mathbf{r_{i}}}\left\{ \bar{C}_{p_{x},\uparrow}\left(\mathbf{r_{i}},\tau\right)\bar{C}_{p_{y},\downarrow}\left(\mathbf{r_{i}},\tau\right)C_{p_{x},\downarrow}\left(\mathbf{r_{i}},\tau\right)C_{p_{y},\uparrow}\left(\mathbf{r_{i}},\tau\right)\right.\right.\\
&+\bar{C}_{p_{y},\uparrow}\left(\mathbf{r_{i}},\tau\right)\bar{C}_{p_{x},\downarrow}\left(\mathbf{r_{i}},\tau\right)C_{p_{x},\downarrow}\left(\mathbf{r_{i}},\tau\right)C_{p_{y},\uparrow}\left(\mathbf{r_{i}},\tau\right)+\bar{C}_{p_{x},\uparrow}\left(\mathbf{r_{i}},\tau\right)\bar{C}_{p_{y},\downarrow}\left(\mathbf{r_{i}},\tau\right)C_{p_{y},\downarrow}\left(\mathbf{r_{i}},\tau\right)C_{p_{x},\uparrow}\left(\mathbf{r_{i}},\tau\right)\\
&\left.+\bar{C}_{p_{y},\uparrow}\left(\mathbf{r_{i}},\tau\right)\bar{C}_{p_{x},\downarrow}\left(\mathbf{r_{i}},\tau\right)C_{p_{y},\downarrow}\left(\mathbf{r_{i}},\tau\right)C_{p_{x},\uparrow}\left(\mathbf{r_{i}},\tau\right)\right\} \bigg )
\end{aligned}
\end{equation*}
\begin{equation}
\begin{aligned}
=&\int\mathcal{D}\bar{C}\mathcal{D}C\prod_{j=1}^{2}\mathcal{D}\bar{\mathit{\Delta}}_{p_{x}p_{y}}^{(j)}\mathcal{D}\mathit{\Delta}_{p_{x}p_{y}}^{(j)}\mathcal{D}\bar{\mathit{\Delta}}_{p_{y}p_{x}}^{(j)}\mathcal{D}\mathit{\Delta}_{p_{y}p_{x}}^{(j)}\exp\bigg(\frac{1}{W}\intop_{0}^{\beta} \mathrm{d}\tau \sum_{\mathbf{r_{i}}}\big\{\bar{\mathit{\Delta}}_{p_{y}p_{x}}^{(2)}\left(\mathbf{r_{i}},\tau\right)\mathit{\Delta}_{p_{y}p_{x}}^{(1)}\left(\mathbf{r_{i}},\tau\right)+\bar{\mathit{\Delta}}_{p_{x}p_{y}}^{(1)}\left(\mathbf{r_{i}},\tau\right)\mathit{\Delta}_{p_{x}p_{y}}^{(2)}\left(\mathbf{r_{i}},\tau\right)\\
&+\bar{\mathit{\Delta}}_{p_{y}p_{x}}^{(1)}\left(\mathbf{r_{i}},\tau\right)\mathit{\Delta}_{p_{x}p_{y}}^{(1)}\left(\mathbf{r_{i}},\tau\right)
+\bar{\mathit{\Delta}}_{p_{x}p_{y}}^{(2)}\left(\mathbf{r_{i}},\tau\right)\mathit{\Delta}_{p_{y}p_{x}}^{(2)}\left(\mathbf{r_{i}},\tau\right)\big \} +\intop_{0}^{\beta} \mathrm{d}\tau \big \{ \left[\mathit{\Delta}_{p_{y}p_{x}}^{(1)}\left(\mathbf{r_{i}},\tau\right)+\mathit{\Delta}_{p_{x}p_{y}}^{(1)}\left(\mathbf{r_{i}},\tau\right)\right]\bar{C}_{p_{x},\uparrow}\left(\mathbf{r_{i}},\tau\right)\bar{C}_{p_{y},\downarrow}\left(\mathbf{r_{i}},\tau\right)\\
&+\left[\mathit{\Delta}_{p_{y}p_{x}}^{(2)}\left(\mathbf{r_{i}},\tau\right)+\mathit{\Delta}_{p_{x}p_{y}}^{(2)}\left(\mathbf{r_{i}},\tau\right)\right]\bar{C}_{p_{y},\uparrow}\left(\mathbf{r_{i}},\tau\right)\bar{C}_{p_{x},\downarrow}\left(\mathbf{r_{i}},\tau\right)+\left[\bar{\mathit{\Delta}}_{p_{x}p_{y}}^{(1)}\left(\mathbf{r_{i}},\tau\right)+\bar{\mathit{\Delta}}_{p_{y}p_{x}}^{(1)}\left(\mathbf{r_{i}},\tau\right)\right]C_{p_{x},\downarrow}\left(\mathbf{r_{i}},\tau\right)C_{p_{y},\uparrow}\left(\mathbf{r_{i}},\tau\right)\\
&+\left[\bar{\mathit{\Delta}}_{p_{x}p_{y}}^{(2)}\left(\mathbf{r_{i}},\tau\right)+\bar{\mathit{\Delta}}_{p_{y}p_{x}}^{(2)}\left(\mathbf{r_{i}},\tau\right)\right]C_{p_{y},\downarrow}\left(\mathbf{r_{i}},\tau\right)C_{p_{x},\uparrow}\left(\mathbf{r_{i}},\tau\right)\big \} \bigg ).
\end{aligned}
\end{equation}
While in the case with $\mu,\nu=p_y, p_z$ and $\mu,\nu=p_x, p_z$, the corresponding expressions can be derived in the same way. We next introduce the following Fourier transformations
\[
C_{\mu,\mathbf{k},n,\sigma}=\frac{1}{\sqrt{\beta N}}\intop_{0}^{\beta}\sum_{\mathbf{r}}C_{\mu,\sigma}\left(\mathbf{r},\tau\right)e^{i\omega_{n}\tau-i\mathbf{k}\cdot\mathbf{r}}\,\mathrm{d}\tau,\qquad\bar{C}_{\mu,\mathbf{k},n,\sigma}=\frac{1}{\sqrt{\beta N}}\intop_{0}^{\beta}\sum_{\mathbf{r}}\bar{C}_{\mu,\sigma}\left(\mathbf{r},\tau\right)e^{-i\omega_{n}\tau+i\mathbf{k}\cdot\mathbf{r}}\,\mathrm{d}\tau,
\]
\begin{equation}
\mathit{\Delta}_{\mu\nu,\mathbf{q},m}^{(j)}=\frac{1}{\sqrt{\beta N}}\intop_{0}^{\beta}\sum_{\mathbf{r}}\mathit{\Delta}_{\mu\nu}^{(j)}\left(\mathbf{r},\tau\right)e^{i\varpi_{m}\tau-i\mathbf{q}\cdot\mathbf{r}}\,\mathrm{d}\tau,\qquad\bar{\mathit{\Delta}}_{\mu\nu,\mathbf{q},m}^{(j)}=\frac{1}{\sqrt{\beta N}}\intop_{0}^{\beta}\sum_{\mathbf{r}}\bar{\mathit{\Delta}}_{\mu\nu}^{(j)}\left(\mathbf{r},\tau\right)e^{-i\varpi_{m}\tau+i\mathbf{q}\cdot\mathbf{r}}\,\mathrm{d}\tau,\label{eq:FTrelations}
\end{equation}
where $N$ is the total number of lattice sites and $\omega_{n}=\left(2n+1\right)n/\beta$,
$\varpi_{m}=2m\pi/\beta$ with $\beta=1/T$
are the fermionic and bosonic Matsubara
frequencies, respectively. Since the superfluid state investigated here
is the BCS-type, we can further approximately rewrite the Hubbard-Stratonovich fields representing the fermionic pairing as
\[
\mathit{\Delta}_{\mu\nu,\mathbf{q},m}^{(j)}=\sqrt{\beta N}\delta_{\mathbf{q},0}\delta_{m,0}\Delta_{\mu\nu}^{(j)},
\]
\begin{equation}
\bar{\mathit{\Delta}}_{\mu\nu,\mathbf{q},m}^{(j)}=\sqrt{\beta N}\delta_{\mathbf{q},0}\delta_{m,0}\bar{\Delta}_{\mu\nu}^{(j)}.\label{eq:SaddlePointAnsatz}
\end{equation}
Then the partition function becomes
\[
\mathcal{Z}=\int\mathcal{D}\left\{ \bar{C},C, \bar{\Delta}^{(j)},\Delta^{(j)} \right\} \,\exp\left(-S_{sp}\left[\bar{C},C,\bar{\Delta}^{(j)},\Delta^{(j)}\right]\right),
\]
\begin{equation}
\begin{aligned}
S_{sp}\left[\bar{C},C,\bar{\Delta},\Delta\right]=&-\frac{N\beta}{W}\sum_{\mu\neq\nu}\left(3\bar{\Delta}_{\mu\mu}^{(1)}\Delta_{\mu\mu}^{(1)}+\bar{\Delta}_{\mu\mu}^{(1)}\Delta_{\nu\nu}^{(1)}\right)-\frac{N\beta}{W}\sum_{\substack{\mu\nu=\left\{ p_{x}p_{y},\right.\\
\left.p_{y}p_{z},p_{z}p_{x}\right\}
}
}\left(\bar{\Delta}_{\mu\nu}^{(1)}\Delta_{\mu\nu}^{(2)}+\bar{\Delta}_{\nu\mu}^{(2)}\Delta_{\nu\mu}^{(1)}+\bar{\Delta}_{\nu\mu}^{(1)}\Delta_{\mu\nu}^{(1)}+\bar{\Delta}_{\mu\nu}^{(2)}\Delta_{\nu\mu}^{(2)}\right) \\
&
+\sum_{\mathbf{k},n,\mu,\nu,\sigma}\bar{C}_{\mu,\mathbf{k},n,\sigma}\left(-i\omega_{n}\delta_{\mu\nu}+\left[\mathscr{H}(\mathbf{k})\right]_{\mu\nu}\right)C_{\nu,\mathbf{k},n,\sigma}-\sum_{\substack{\mathbf{k},n\\
\mu\neq\nu
}
}\left(\left[3\bar{\Delta}_{\mu\mu}^{(1)}+\bar{\Delta}_{\nu\nu}^{(1)}\right]C_{\mu,\mathbf{-k},-n,\downarrow}C_{\mu,\mathbf{k},n,\uparrow}\right. \\
&
\left.+\left[3\Delta_{\mu\mu}^{(1)}+\Delta_{\nu\nu}^{(1)}\right]\bar{C}_{\mu,\mathbf{k},n,\uparrow}\bar{C}_{\mu,\mathbf{-k},-n,\downarrow}\right)-\sum_{\substack{\mathbf{k},n\\
\mu\nu=\left\{ p_{x}p_{y},\right.\\
\left.p_{y}p_{z},p_{z}p_{x}\right\}
}
}\left(\left[\bar{\Delta}_{\mu\nu}^{(1)}+\bar{\Delta}_{\nu\mu}^{(1)}\right]C_{\mu,\mathbf{-k},-n,\downarrow}C_{\nu,\mathbf{k},n,\uparrow}\right.\\
&\left.+\left[\Delta_{\mu\nu}^{(1)}+\Delta_{\nu\mu}^{(1)}\right]\bar{C}_{\mu,\mathbf{k},n,\uparrow}\bar{C}_{\nu,\mathbf{-k},-n,\downarrow}\right)
-\sum_{\substack{\mathbf{k},n\\
\mu\nu=\left\{ p_{y}p_{x},\right.\\
\left.p_{z}p_{y},p_{x}p_{z}\right\}
}
}\left(\left[\bar{\Delta}_{\mu\nu}^{(2)}+\bar{\Delta}_{\nu\mu}^{(2)}\right]C_{\mu,\mathbf{-k},-n,\downarrow}C_{\nu,\mathbf{k},n,\uparrow}\right.\\
&\left.+\left[\Delta_{\mu\nu}^{(2)}+\Delta_{\nu\mu}^{(2)}\right]\bar{C}_{\mu,\mathbf{k},n,\uparrow}\bar{C}_{\nu,\mathbf{-k},-n,\downarrow}\right).\label{eq:SaddlePointPFIntermediate}
\end{aligned}
\end{equation}
We then further represent the action in Nambu representation
\begin{equation}
\begin{aligned}
S_{sp}\left[\bar{C},C,\bar{\Delta},\Delta\right]=&-\frac{N\beta}{W}\sum_{\mu\neq\nu}\left(3\bar{\Delta}_{\mu\mu}^{(1)}\Delta_{\mu\mu}^{(1)}+\bar{\Delta}_{\mu\mu}^{(1)}\Delta_{\nu\nu}^{(1)}\right)-\frac{N\beta}{W}\sum_{\substack{\mu\nu=\left\{ p_{x}p_{y},\right.\\
\left.p_{y}p_{z},p_{z}p_{x}\right\}
}
}\left(\bar{\Delta}_{\mu\nu}^{(1)}\Delta_{\mu\nu}^{(2)}+\bar{\Delta}_{\nu\mu}^{(2)}\Delta_{\nu\mu}^{(1)}+\bar{\Delta}_{\nu\mu}^{(1)}\Delta_{\mu\nu}^{(1)}+\bar{\Delta}_{\mu\nu}^{(2)}\Delta_{\nu\mu}^{(2)}\right)\\
&+\sum_{\mathbf{k},n}\bar{\eta}_{\mathbf{k},n}\left(-i\omega_{n}\mathbb{I}_{6}+H_{BdG}\left(\mathbf{k}\right)\right)\eta_{\mathbf{k},n},\label{eq:NambuSpaceAction}
\end{aligned}
\end{equation}
where $\eta_{\mathbf{k},n}=\left(\begin{array}{cccccc}
C_{p_{x},\mathbf{k},n,\uparrow} & C_{p_{y},\mathbf{k},n,\uparrow} & C_{p_{z},\mathbf{k},n,\uparrow} & \bar{C}_{p_{x},\mathbf{-k},-n,\downarrow} & \bar{C}_{p_{y},\mathbf{-k},-n,\downarrow} & \bar{C}_{p_{z},\mathbf{-k},-n,\downarrow}\end{array}\right)^{T}$ is the Nambu spinor.
$\mathbb{I}_{6}$ is the $6\times6$ unit matrix and $H_{BdG}\left(\mathbf{k}\right)$
is the $6\times6$ Bogoliubov-de Gennes matrix
\[
H_{BdG}\left(\mathbf{k}\right)=\left(\begin{array}{cc}
\mathscr{H}(\mathbf{k}) & -M\\
-\overline{M} & -\mathscr{H}^{T}(-\mathbf{k})
\end{array}\right),
\]
\[
M=\left(\begin{array}{ccc}
3\Delta_{p_{x}p_{x}}^{(1)}+\Delta_{p_{y}p_{y}}^{(1)}+\Delta_{p_{z}p_{z}}^{(1)} & \Delta_{p_{x}p_{y}}^{(1)}+\Delta_{p_{y}p_{x}}^{(1)} & \Delta_{p_{x}p_{z}}^{(2)}+\Delta_{p_{z}p_{x}}^{(2)}\\
\Delta_{p_{x}p_{y}}^{(2)}+\Delta_{p_{y}p_{x}}^{(2)} & 3\Delta_{p_{y}p_{y}}^{(1)}+\Delta_{p_{x}p_{x}}^{(1)}+\Delta_{p_{z}p_{z}}^{(1)} & \Delta_{p_{y}p_{z}}^{(1)}+\Delta_{p_{z}p_{y}}^{(1)}\\
\Delta_{p_{x}p_{z}}^{(1)}+\Delta_{p_{z}p_{x}}^{(1)} & \Delta_{p_{y}p_{z}}^{(2)}+\Delta_{p_{z}p_{y}}^{(2)} & 3\Delta_{p_{z}p_{z}}^{(1)}+\Delta_{p_{x}p_{x}}^{(1)}+\Delta_{p_{y}p_{y}}^{(1)}
\end{array}\right),
\]
with \begin{equation}
\overline{M}=\left(\begin{array}{ccc}
3\bar{\Delta}_{p_{x}p_{x}}^{(1)}+\bar{\Delta}_{p_{y}p_{y}}^{(1)}+\bar{\Delta}_{p_{z}p_{z}}^{(1)} & \bar{\Delta}_{p_{x}p_{y}}^{(1)}+\bar{\Delta}_{p_{y}p_{x}}^{(1)} & \bar{\Delta}_{p_{x}p_{z}}^{(2)}+\bar{\Delta}_{p_{z}p_{x}}^{(2)}\\
\bar{\Delta}_{p_{x}p_{y}}^{(2)}+\bar{\Delta}_{p_{y}p_{x}}^{(2)} & 3\bar{\Delta}_{p_{y}p_{y}}^{(1)}+\bar{\Delta}_{p_{x}p_{x}}^{(1)}+\bar{\Delta}_{p_{z}p_{z}}^{(1)} & \bar{\Delta}_{p_{y}p_{z}}^{(1)}+\bar{\Delta}_{p_{z}p_{y}}^{(1)}\\
\bar{\Delta}_{p_{x}p_{z}}^{(1)}+\bar{\Delta}_{p_{z}p_{x}}^{(1)} & \bar{\Delta}_{p_{y}p_{z}}^{(2)}+\bar{\Delta}_{p_{z}p_{y}}^{(2)} & 3\bar{\Delta}_{p_{z}p_{z}}^{(1)}+\bar{\Delta}_{p_{x}p_{x}}^{(1)}+\bar{\Delta}_{p_{z}p_{z}}^{(1)}
\end{array}\right).\label{eq:BdGHamiltonian}
\end{equation}
Integrating out the fermionic fields, the partition function can be expressed as
\begin{equation}
\mathcal{Z}=\exp\left(-\beta NF_{sp}(\bar{\Delta},\Delta,T,\mu)\right),
\end{equation}
\begin{equation}
\begin{aligned}
F_{sp}(\bar{\Delta},\Delta,T,\mu)=&-\frac{1}{W}\sum_{\mu\neq\nu}\left(3\bar{\Delta}_{\mu\mu}^{(1)}\Delta_{\mu\mu}^{(1)}+\bar{\Delta}_{\mu\mu}^{(1)}\Delta_{\nu\nu}^{(1)}\right)-\frac{1}{W}\sum_{\substack{\mu\nu=\left\{ p_{x}p_{y},\right.\\
\left.p_{y}p_{z},p_{z}p_{x}\right\}
}
}\left(\bar{\Delta}_{\mu\nu}^{(1)}\Delta_{\mu\nu}^{(2)}+\bar{\Delta}_{\nu\mu}^{(2)}\Delta_{\nu\mu}^{(1)}+\bar{\Delta}_{\nu\mu}^{(1)}\Delta_{\mu\nu}^{(1)}+\bar{\Delta}_{\mu\nu}^{(2)}\Delta_{\nu\mu}^{(2)}\right)\\
&-\frac{1}{N\beta}\sum_{\mathbf{k},n}\ln\left\{ -\det\left[-i\omega_{n}\mathbb{I}_{6}+H_{BdG}\left(\mathbf{k}\right)\right]\right\} .\label{eq:SaddlePointPF}
\end{aligned}
\end{equation}
Then the filling of the system and the saddle values of the Hubbard-Stratonovich fields representing the fermionic pairing can be determined by
\begin{equation}
n=-\frac{\partial F_{sp}(\bar{\Delta},\Delta,T,\mu)}{\partial\mu},\qquad\frac{\partial F_{sp}(\bar{\Delta},\Delta,T,\mu)}{\partial\bar{\Delta}_{\mu\nu}^{(j)}}=0,\qquad\frac{\partial F_{sp}(\bar{\Delta},\Delta,T,\mu)}{\partial\Delta_{\mu\nu}^{(j)}}=0.\label{eq:SelfConsistentConditions}
\end{equation}
From Eq.~\eqref{eq:SelfConsistentConditions}, we can define the pairing order parameters
as $\Delta_{\mu\nu}\equiv\Delta_{\mu\nu}^{(1)},\Delta_{\mu\nu}^{*}\equiv\bar{\Delta}_{\mu\nu}^{(1)}$
when $\mu=\nu$, and for $\mu\neq\nu$,
$\Delta_{\mu\nu}\equiv\Delta_{\mu\nu}^{(1)}=\Delta_{\mu\nu}^{(2)},\Delta_{\mu\nu}^{*}\equiv\bar{\Delta}_{\mu\nu}^{(1)}=\bar{\Delta}_{\mu\nu}^{(2)}$.
Therefore, we can rewrite the $H_{BdG}\left(\mathbf{k}\right)$ as
\[
H_{BdG}\left(\mathbf{k}\right)=\left(\begin{array}{cc}
\mathscr{H}(\mathbf{k}) & -M\\
-\left(M^{T}\right)^{*} & -\mathscr{H}^{T}(-\mathbf{k})
\end{array}\right),
\]
\begin{equation}
M=\left(\begin{array}{ccc}
3\Delta_{p_{x}p_{x}}+\Delta_{p_{y}p_{y}}+\Delta_{p_{z}p_{z}} & \Delta_{p_{x}p_{y}}+\Delta_{p_{y}p_{x}} & \Delta_{p_{x}p_{z}}+\Delta_{p_{z}p_{x}}\\
\Delta_{p_{x}p_{y}}+\Delta_{p_{y}p_{x}} & 3\Delta_{p_{y}p_{y}}+\Delta_{p_{x}p_{x}}+\Delta_{p_{z}p_{z}} & \Delta_{p_{y}p_{z}}+\Delta_{p_{z}p_{y}}\\
\Delta_{p_{x}p_{z}}+\Delta_{p_{z}p_{x}} & \Delta_{p_{y}p_{z}}+\Delta_{p_{z}p_{y}} & 3\Delta_{p_{z}p_{z}}+\Delta_{p_{x}p_{x}}+\Delta_{p_{y}p_{y}}
\end{array}\right).\label{eq:MeanFieldBdG}
\end{equation}

{\section{Berezinskii-Kosterlitz-Thouless (BKT) transition}\label{sec:tbkt}
It is well known that at finite temperature the superfluidity
of $2$D atomic Fermi gases is characterized by the vortex-antivortex
binding. The relevant mechanism is the Berezinskii-Kosterlitz-Thouless (BKT) ~\cite{1972_Berezinskii,1973_KosterlitzThouless} transition occurring at a characteristic temperature $T_{BKT}$. The BKT transition in $2$D is associated with the spontaneous vortex formation. A unique feature of such a
transition is an universal jump in the superfluid density~\cite{1977_NelsonKosterlitz}. To further determine the superfluid
density, we imposing a phase twist (e.g. see ~\cite{2009_Paananen,BKTPhaseTransition2D,2012_Yanay_Mueller_preprint}), i.e., a supercurrent, on the order
parameter as
\begin{equation}
\Delta_{\mu\nu}\rightarrow\Delta_{\mu\nu}\exp\left(i\cdot2\mathbf{\Theta \cdot r_i}\right)=\Delta_{\mu\nu}\exp\left(i\frac{2\Theta_{x}n_{x}}{N_x} +i\frac{2\Theta_{y} n_{y}}{N_y}\right),\label{eq:ImposingPhase}
\end{equation}
where $\mathbf{r_i}=n_{x}a\vec{e}_{x}+n_{y}a\vec{e}_{y}$ labels the
lattice site. $\mathbf{\Theta}$ captures the linear phase variation on the order
parameter and $N_{x}$, $N_{y}$ represents the site number along $x$ and $y$ direction, respectively. This imposed phase gradient gives the system a kinetic energy, which corresponds
to the free energy difference $\triangle F \equiv F_{\bf \Theta}-F_0$, where $F_{\bf \Theta}$ is the free energy within the phase variation and
$F_0$ is the free energy without the phase variation. We then approximate $\triangle F$ up to the second order in $\mathbf{\Theta}$ as
$\triangle F\simeq\sum_{\alpha,\beta=x,y}F_{\alpha\beta}^{(2)}\Theta_{\alpha}\Theta_{\beta}
$. Due to the reflection symmetry of the system, when
$\alpha\neq\beta$, $F_{\alpha\beta}^{(2)}$ vanishes and we further have
\begin{eqnarray}
\begin{aligned}
F_{\alpha\alpha}^{(2)}=& \frac{T}{2N}\sum_{\mathbf{k},n,\mu,\nu}\left[\frac{\partial^{2}\mathscr{H}\left(\mathbf{k}\right)}{\partial^{2}k_{\alpha}}\right]_{\mu\nu}\left(-\left[G_{\mathbf{k},n,\uparrow}\right]_{\nu\mu}-\left[G_{\mathbf{-k},-n,\downarrow}\right]_{\mu\nu}\right)\\
 &+\frac{T}{2}\sum_{\mathbf{k},n,\mu,\nu,\mu',\nu'}\left[\frac{\partial \mathscr{H}\left(\mathbf{k}\right)}{\partial k_{\alpha}}\right]_{\mu\nu}\left[\frac{\partial \mathscr{H}\left(\mathbf{k}\right)}{\partial k_{\alpha}}\right]_{\mu'\nu'}\left(\left[G_{\mathbf{k},n,\uparrow}\right]_{\nu\mu'}\left[G_{\mathbf{k},n,\uparrow}\right]_{\nu'\mu}+\left[G_{\mathbf{k},n,\downarrow}\right]_{\nu\mu'}\left[G_{\mathbf{k},n,\downarrow}\right]_{\nu'\mu}\right.\\
 &\left.+\left[F_{\mathbf{k},n}\right]_{\nu\mu'}\left[F_{\mathbf{k},n}^{\dagger}\right]_{\nu'\mu}+\left[F_{\mathbf{k},n}\right]_{\nu'\mu}\left[F_{\mathbf{k},n}^{\dagger}\right]_{\nu\mu'}\right),\label{eq:SuperfluidDensity}
 \end{aligned}
\end{eqnarray}
with $\left[G_{\mathbf{k},n,\sigma}\right]_{\mu\nu}=\left\langle C_{\mu,\mathbf{k},n,\sigma}\bar{C}_{\nu,\mathbf{k},n,\sigma}\right\rangle $ and
$\left[F_{\mathbf{k},n}\right]_{\mu\nu}=\left\langle C_{\mu,\mathbf{k},n,\uparrow}C_{\nu,\mathbf{-k},-n,\downarrow}\right\rangle$ are matrix elements of the normal and anomalous Greens's functions. Then the BKT transition temperature can be determined through $k_{B}T_{BKT}=\frac{\pi}{4} F^{(2)}$ with $F^{(2)}\equiv (F_{xx}^{(2)}+F_{yy}^{(2)})/2$.

\end{document}